\documentclass[letterpaper, 10 pt, conference]{ieeeconf}  

\IEEEoverridecommandlockouts                              

\usepackage{cite}
\usepackage{amsmath,amssymb,amsfonts}
\usepackage{algorithmic}
\usepackage[graphicx]{realboxes}
\usepackage{textcomp}
\usepackage[table,xcdraw]{xcolor}
\usepackage{multirow}
\usepackage{array}
\usepackage{tikz}
\usepackage{siunitx}
\usepackage{lscape}
\usepackage{adjustbox}
\usepackage{tabularx}
\usepackage{mleftright}
\usepackage{etoolbox}
\usepackage{xcolor}
\usepackage{amsmath}
\usepackage{amssymb}
\usepackage[normalem]{ulem}
\usepackage{tablefootnote}
\usepackage{verbatim}
\usepackage{graphicx}
\usepackage{booktabs}

\title{\LARGE \bf
Towards Natural Brain-Machine Interaction using Endogenous Potentials based on Deep Neural Networks}

\author{Hyung-Ju Ahn$^{1}$, Dae-Hyeok Lee$^{1}$, Ji-Hoon Jeong$^{1}$, and Seong-Whan Lee$^{2}$
\thanks{*This research was supported by the Defense Challengeable Future Technology Program of Agency for Defense Development, Republic of Korea.}
\thanks{$^{1}$H.-J. Ahn, D.-H. Lee, and J.-H. Jeong are with the Department of Brain and Cognitive Engineering, Korea University, Anam-dong, Seongbuk-ku, Seoul 02841, Korea.
        {\tt\small hj\_ahn@korea.ac.kr, lee\_dh@korea.ac.kr, and jh\_jeong@korea.ac.kr}}%
\thanks{$^{2}$S.-W. Lee is with the Department of Artificial Intelligence, Korea University, Anam-dong, Seongbuk-ku, Seoul 02841, Korea.
        {\tt\small sw.lee@korea.ac.kr}}%
}

\begin{document}

\maketitle
\thispagestyle{empty}
\pagestyle{empty}

\begin{abstract}
Human-robot collaboration has the potential to maximize the efficiency of the operation of autonomous robots. Brain-machine interface (BMI) would be a desirable technology to collaborate with robots since the user's intention or state of mind can be translated from the neural activities. However, the electroencephalogram (EEG), which is one of the most popularly used non-invasive BMI modalities, has low accuracy and a limited degree of freedom (DoF) due to a low signal-to-noise ratio. Thus, improving the performance of multi-class EEG classification is crucial to develop more flexible BMI-based human-robot collaboration. In this study, we investigated the possibility for inter-paradigm classification of multiple endogenous BMI paradigms, such as motor imagery (MI), visual imagery (VI), and speech imagery (SI), to enhance the limited DoF while maintaining robust accuracy. We conducted the statistical and neurophysiological analyses on MI, VI, and SI and classified three paradigms using the proposed temporal information-based neural network (TINN). We confirmed that statistically significant features could be extracted on different brain regions when classifying three endogenous paradigms. Moreover, our proposed TINN showed the highest accuracy of 0.93 compared to the previous methods for classifying three different types of mental imagery tasks (MI, VI, and SI).\\
\end{abstract}
\section{INTRODUCTION}

Brain-machine interface (BMI) is a promising technology to control the peripheral devices using intrinsic brain activity. BMI has been developed to rehabilitate neurological impairment patients, estimate one's mental state\cite{lee2020continuous}, or enhance human's ability by using invasive or non-invasive ways\cite{prichard2014effects}. Other than clinical and rehabilitation usages of traditional BMI, the practical BMI applications for able-bodied users are considered as an emerging trend\cite{jantz2017brain}. One of them is a collaboration between humans and autonomous robots.

Recently, autonomous robots are being vigorously developed for industrial, commercial, and military purposes. They are not only coping well with repetitive tasks and high payloads at high speeds but also can be utilized in hazardous environments without casualty risks such as disaster responses and battlefields. However, their adaptability and resilience to nonlinear environments are far less than humans. Thus, high productivity and adaptability would be achieved by a joint collaboration of humans and robots. Meanwhile, the restrictions prevent close collaboration to protect human operators due to current systems of safety, which are being primarily reactive. Therefore, recognizing the users' intention through BMI would be desirable for human-robot collaboration. Moreover, BMI-based human-robot collaboration has the potential of natural cooperation between humans and robots with hands-free teleoperation.

Electroencephalogram (EEG) is one of the most actively used non-invasive BMI modalities. EEG systems have a relatively lower equipment price than other modalities, and their signals have a relatively high temporal resolution \cite{lee2020frontal}. Various kinds of EEG-based BMI paradigms have been developed to induce particular brain signals in a specific condition. Exogenous BMI paradigms such as P300\cite{fazel2012p300}, event-related potential\cite{luck2000event,lee2018high}, or steady-state visual evoked potential\cite{liu2014recent} use visual cues to induce the user's responses. More noticeable features can be extracted from EEG signals with external stimulation. However, in terms of the practical BMI system, dependency on external stimuli can hinder users' multi-tasking ability. Moreover, watching the flickering lights induces fatigue to users and even seizures for some epilepsy patients\cite{fisher2005photic}.
On the other hand, endogenous BMI paradigms, including motor imagery (MI), visual imagery (VI), or speech imagery (SI), do not require any external stimulation to use the system. Endogenous BMI paradigms are described as imagining a movement, a visual image, or a word's pronunciation. The independence from the external stimuli would provide more intuitive control for the BMI applications since the EEG signals will be classified on users' imaginations\cite{suk2011subject}. 

Lately, the EEG-based human-robot collaboration is being actively explored. Buerkle \textit{et al}.\cite{buerkle2021eeg} proposed the adaptive speed and torque control system for cooperative robotic arm based on detecting user's movement intention. The decoding performance of right hand and left-hand movement intention was derived 0.5 s before their execution with the accuracy of 0.84--0.92. Liu \textit{et al}.\cite{liu2021brain} developed the MI-based hands-free teleoperation of construction robots with 0.9 accuracy of 3-class classification. 
However, EEG-based endogenous paradigms have a trade-off relationship between classification accuracy and DoF due to the low signal-to-noise ratio. In addition, each type of sensory imagery is mapped locally into each brain region. Hence, it is hard to classify multiple intentions with the same types of sensory imagery. Thus, the use of multiple endogenous paradigms has the possibility to enhance the limited DoF while maintaining high accuracy.

Although many kinds of research have been made for each of the endogenous paradigms, only a few studies have been researched on multiple endogenous paradigms. Koizumi \textit{et al}.\cite{koizumi2018development} investigated that high-$\gamma$ frequency band power and EEG signals from the prefrontal cortex are valuable to classify both in VI and SI paradigms. Nevertheless, electromyogram artifacts consisting of high-frequency components contaminate EEG in this frequency range, and it is still unclear that EEG signals from the prefrontal cortex are the most useful to classify imagined speech tasks. Chholak \textit{et al}.\cite{chholak2019visual} conducted a magnetoencephalographic experiment to distinguish different brain activation patterns between the MI and VI tasks. They confirmed different event-related synchronization/desynchronization of $\alpha$- and $\beta$-wave activity and different evoked responses in the frontal cortex during the MI and VI tasks.

\begin{figure}[!]
\scriptsize
\centerline{\includegraphics[width=\columnwidth]{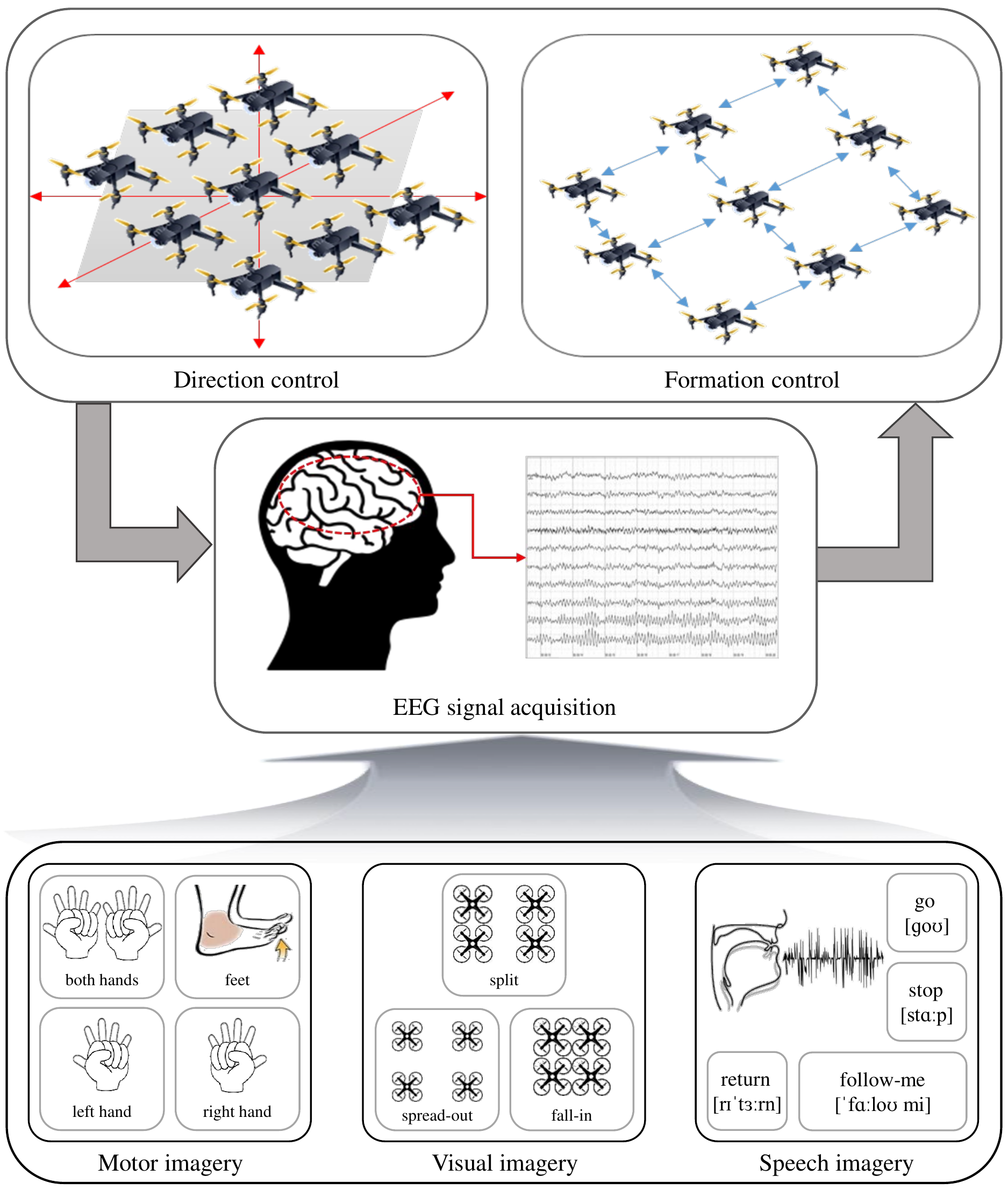}}
\caption{The conceptual system architecture for collaborative control of BMI-based semi-autonomous drone swarm. The figure shows the commands for direction and formation control of drone swarm and details of subclasses in each endogenous BMI paradigm.}
\end{figure}

Recently, numerous deep learning algorithms have been designed to classify EEG signals. A convolutional neural network (CNN) has been successfully applied to EEG-based BMIs for end-to-end feature extraction and classification\cite{lee2020classification}, as well as computer vision and speech recognition. Schirrmeister \textit{et al}.\cite{schirrmeister2017deep} proposed CNN architectures to decode raw EEG signals with a range of different architectures. Although the features were not fixed priors, they achieved as good performance as filter bank common spatial pattern (FBCSP)\cite{ang2008filter}, which has been most popularly applied to feature extraction for MI classification using EEG. Lawhern \textit{et al}.\cite{lawhern2018eegnet} designed a single CNN architecture that is robust enough to classify EEG signals from different BMI paradigms (P300 visual-evoked potentials, movement-related cortical potentials\cite{jeong2020decoding}, error-related negativity responses\cite{yeung2004neural}, and sensory-motor rhythms).

In this study, we designed the experimental protocols to potentially collaborate with drone swarm using three endogenous BMI paradigms. In order to validate our hypothesis on the robust multi-class EEG classification by using different paradigms, we conducted the statistical and neurophysiological analyses to investigate the meaningful features for classifying MI, VI, and SI paradigms. Also, we classified three paradigms with high and robust accuracy (0.93) by using our proposed temporal information-based neural network (TINN). To the best of our knowledge, this is the first attempt to classify all three MI, VI, and SI paradigms on the endogenous BMI, which has the potential of expanding the DoF for BMI through inter-paradigm classification. In this study, Fig. 1 describes the conceptual system architecture of controlling the drone swarms utilizing the endogenous BMI paradigms. Fig.1 corresponds to subclasses in each paradigm, which we devised to control the flight mission of drone swarm with various directions and formations. The BMI-based human-robot collaboration would be valuable for multiplying the workforces in diverse fields, and it has the potential of enhancing the multi-tasking ability of humans. It would be helpful for handling the time-critical cases, such as activating the cyber defense system or search and rescue mission control.

\section{MATERIALS AND METHODS}

\subsection{Participants}
Ten healthy subjects (S1-S10, ten males, aged 25.5 ($\pm 3.1$)) participated in the experiments. The Institutional Review Board at Korea University approved all experiments (KUIRB-2020-0318-01). Before the experiment, we informed them to get adequate sleep (over seven hours) and avoid any alcohol the day before. Subjects were informed about the experimental protocols and procedures. They provided their written consent according to the Declaration of Helsinki.
\begin{table}[!]
\caption{Details of the TINN structure. Each block is CNN and BiLSTM, repectively.}
\renewcommand{\arraystretch}{1.0}
\scriptsize
\resizebox{\columnwidth}{!}
{%
\begin{tabular}{ccccc}
\hline
Block               & Layer           & Output               & Kernel        & Stride       \\ \hline
\multirow{3}{*}{I}  & Convolution     & {[}1, 25, 64, 327{]} & 1$\times$50 & 1$\times$1 \\
                    & Convolution     & {[}1, 25, 1, 327{]} & 64$\times$1 & 1$\times$1 \\
                    & Average pooling & {[}1, 25, 1, 40{]}   & 1$\times$8  & 1$\times$8 \\
                    & Convolution     & {[}1, 50, 1, 33{]} & 1$\times$8 & 1$\times$8 \\
                    & Average pooling & {[}1, 50, 1, 4{]}   & 1$\times$8  & 1$\times$8 \\ \hline
\multirow{3}{*}{II} & BiLSTM          & {[}1, 4, 200{]}     & -  & - \\
                    & Average pooling & {[}1, 4, 25{]}      & 1$\times$8  & 1$\times$8 \\
                    & Dense           & {[}1, 3{]}           & -             & -            \\ \hline
\end{tabular}}
\end{table}

\begin{table*}[!]
\caption{Statistical comparison of PSD among endogenous BMI paradigms in different frequency bands}
\renewcommand{\arraystretch}{1.5}
\scriptsize
\resizebox{\textwidth}{!}
{%
\begin{tabular}{ccccccccccccc}
\hline
\multirow{2}{*}{Factor} & \multicolumn{2}{c}{$\delta$-band} & \multicolumn{2}{c}{$\theta$-band} & \multicolumn{2}{c}{$\alpha$-band} & \multicolumn{2}{c}{$\beta$-band} & \multicolumn{2}{c}{Low $\gamma$-band} & \multicolumn{2}{c}{High $\gamma$-band} \\ \cline{2-13} 
                        & \textit{F}-value        & \textit{p}-value       & \textit{F}-value        & \textit{p}-value       & \textit{F}-value        & \textit{p}-value       & \textit{F}-value       & \textit{p}-value       & \textit{F}-value          & \textit{p}-value         & \textit{F}-value          & \textit{p}-value          \\ \hline 
Paradigms               & 46.57          & $<$0.001       & 40.69           & $<$0.001       & 12.05           & $<$0.001       & 11.45         & $<$0.001       & 19.09            & $<$0.001         & 32.16              & $<$0.001          \\ \hline
\end{tabular}}
\end{table*}
\begin{figure*}[!]
\centering
\scriptsize
\centerline{\includegraphics[width=\textwidth]{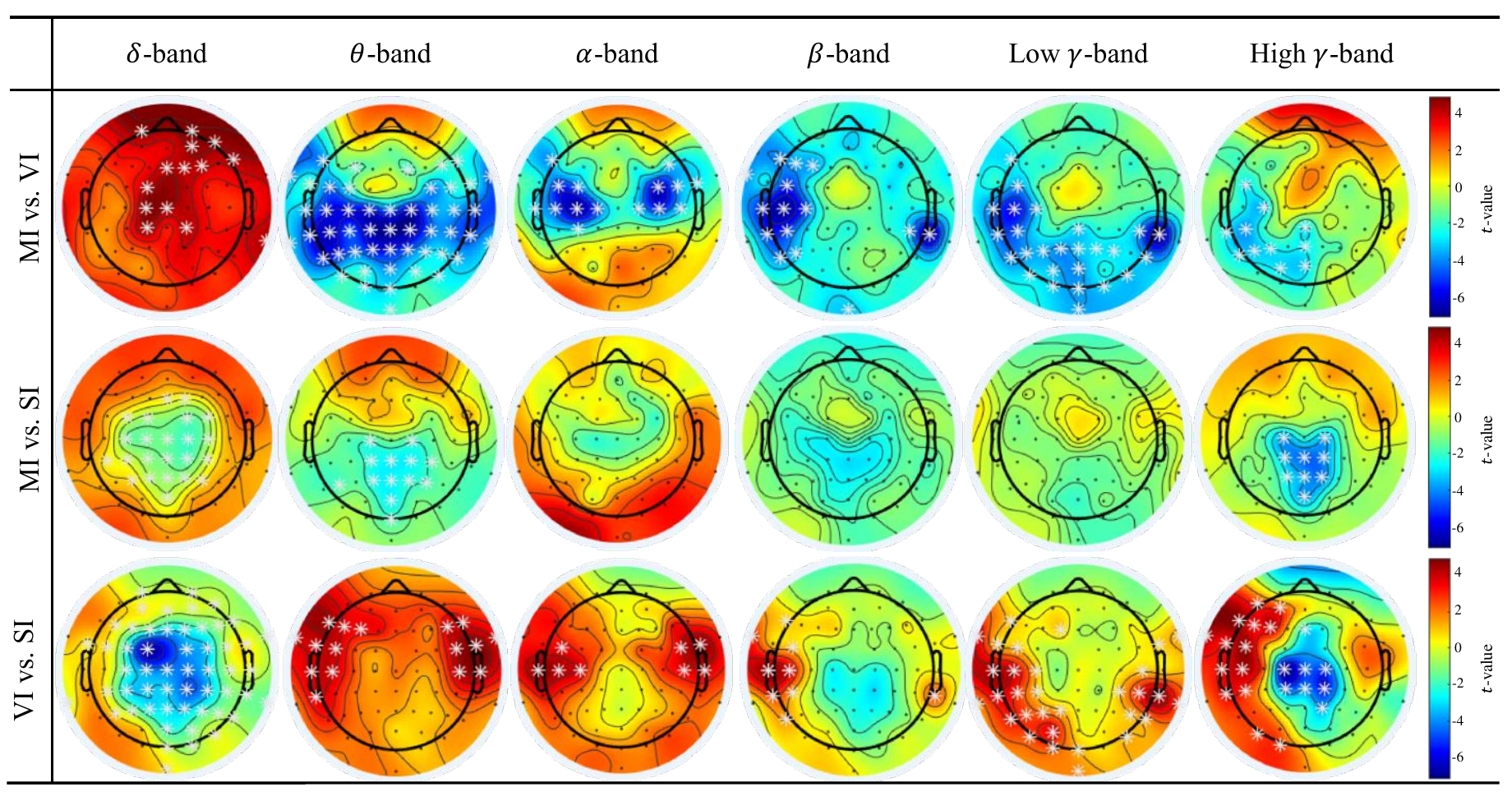}}
\caption{Differences in spectral power between different paradigms and frequency ranges. The statistical results represent \textit{t}-values in each frequency band for differences between two different endogenous BMI paradigms. The blue color reflects relatively lower activity, whereas the red color reflects higher activity compared to another paradigm. The white asterisk indicates a significant electrode in spectral power (\textit{p}$<$0.001 with Bonferroni's correction).}
\end{figure*}

\subsection{EEG Signal Acquisition}
We acquired the EEG data with MI, VI, and SI experiments using our previous paradigms \cite{lee2020design}. We measured the subjects' EEG signals using BrainAmp (BrainProduct GmbH, Germany) and 64 Ag/AgCl electrodes according to the 10/20 international system. The reference electrode was placed at FCz, and the ground electrode was placed at FPz. We set up the sampling rate to 500 Hz, and a 60 Hz notch filter was applied. Before the acquisition of EEG data, all electrodes' impedance was kept below 10 k$\Omega$ by injecting conductive gel.

\subsection{Data Description}
Figure 1. shows specific imagery tasks in each paradigm. The MI tasks are kinesthetic movement imagery of left, right, and both hand, and feet. For the VI tasks, subjects were instructed to visually imagine the scene of three different swarm behaviors (split, spread-out, and fall-in) after watching the video clips of corresponding tasks. In the case of the SI tasks, four kinds of words were presented to imagine the given words' pronunciation (go, stop, follow-me, and return). 

Also, the duration for imagery tasks was set differently for each paradigm. Imagery duration for the MI and VI tasks was set to 4 s since the imagery of kinesthetic movement or visual representation can be continued for few seconds in each trial. However, pronunciation of each word only takes 1 or 2 s for the SI tasks. Imagery duration for the SI tasks was set to 1.5 s and repeated consecutively four times in a single run to match the data length of training dataset. Each of the tasks was repeated for 50 trials. In total, we acquired 200 trials for MI, 150 trials for VI, and 800 trials for SI. Further details of the experimental protocols are included in the preceding paper\cite{lee2020design}. 

\subsection{Data Preprocessing}
EEG data were pre-processed using the BBCI Toolbox\cite{blankertz2010berlin}. The data were downsampled to 250 Hz and bandpass filtered with a fifth-order Butterworth filter in the frequency range of 0.5 to 120 Hz. In MI and VI dataset, we used the sliding window technique with 1.5 s of the time window and 0.7 s of overlapping to divide 4 s of one epoch into 1.5 s of four epochs. Labels of tasks were concatenated into a single class for each paradigm to analyze the factors that could affect the classification of different endogenous paradigms. Consequently, the dataset consisted of 800 epochs for MI, 600 epochs for VI, and 800 epochs for SI, with 1.5 s of duration. Furthermore, 600 epochs out of 800 epochs were randomly selected for each of MI and SI data to standardize the number of training dataset for paradigm classification.

\begin{table}[!]
\caption{The statistical comparison of PSD among endogenous BMI paradigms with the combination of specific classes}
\renewcommand{\arraystretch}{1.1}
\Large
\resizebox{\columnwidth}{!}
{%
\begin{tabular}{ccc|ccc}
\hline
Group                  & \textit{F}-value & \textit{p}-value & Group                   & \textit{F}-value & \textit{p}-value \\ \hline
left-split-go          & 25.93   & $>$0.05    & right-spread out-return & 22.87   & $>$0.05    \\
left-split-stop        & 26.03   & $<$\textbf{0.05}   & right-fall in-go        & 23.55   & $>$0.05    \\
left-split-return      & 25.51   & $<$\textbf{0.05}   & right-fall in-stop      & 23.88   & $<$\textbf{0.05}   \\
left-spread out-go     & 32.83   & $>$0.05    & right-fall in-return    & 23.39   & $<$\textbf{0.05}   \\
left-spread out-stop   & 33.09   & $>$0.05    & feet-split-go           & 22.25   & $>$0.05    \\
left-spread out-return & 32.42   & $>$0.05    & feet-split-stop         & 22.37   & $<$\textbf{0.05}   \\
left-fall in-go        & 33.46   & $>$0.05    & feet-split-return       & 21.92   & $<$\textbf{0.05}   \\
left-fall in-stop      & 33.73   & $<$\textbf{0.05}   & feet-spread out-go      & 28.84   & $>$0.05    \\
left-fall in-return    & 33.04   & $<$\textbf{0.05}   & feet-spread out-stop    & 29.12   & $>$0.05    \\
right-split-go         & 16.97   & $>$0.05    & feet-spread out-return  & 28.53   & $>$0.05    \\
right-split-stop       & 17.13   & $<$\textbf{0.05}   & feet-fall in-go         & 29.43   & $>$0.05    \\
right-split-return     & 16.79   & $<$\textbf{0.05}   & feet-fall in-stop       & 29.73   & $<$\textbf{0.05}   \\
right-spread out-go    & 23.02   & $>$0.05    & feet-fall in-return     & 29.12   & $<$\textbf{0.05}   \\
right-spread out-stop  & 23.34   & $>$0.05    &                         &         &         \\ \hline

\end{tabular}}
\end{table}

\subsection{Data Analysis}
EEGLAB Toolbox\cite{delorme2004eeglab} (version 2021.0) and BBCI toolbox \cite{blankertz2010berlin} were used for data analyses. Due to a small number of samples, Shapiro-Wilk test and Levene's test were initially performed to validate normality and homoscedasticity. To investigate the difference among MI, VI, and SI tasks for various frequency bands, two‐way analysis of variance (ANOVA, at a significance level of \textit{p}$<$0.001) was applied on the spectral power. One factor was the paradigm and the other factor was the channel (spatial information).
Six frequency bands ($\delta$-: 0.5-4 Hz, $\theta$-: 4-8 Hz, $\alpha$-: 8-14 Hz, $\beta$-: 14-30 Hz, low $\gamma$-: 30-60 Hz, and high $\gamma$-band: 60-120 Hz) were selected to extract the power spectral density (PSD), respectively. The statistical comparisons of PSD were performed by paired \textit{t}‐test with Bonferroni's correction for post-hoc analysis. EEGLAB toolbox was used to plot the scalp topographies of paired \textit{t}-test results. Event-related spectral perturbation (ERSP) analyses were performed between 0.5 and 120 Hz to measure the variation of spectral power of MI, VI, and SI. We set the baseline from 0 to 200 ms after the start of the imagination.

\subsection{Temporal information-based neural network (TINN)}
As a decoding method, we proposed the temporal information-based neural network (TINN). The TINN took the form of a hybrid deep learning model using the CNN and BiLSTM frameworks. CNN extracts spatial features and temporal information of brain activities, and BiLSTM fortifies the temporal features. 
Table I shows the architecture of our proposed TINN. During the training, the input data with a size of 64$\times$376 (channel$\times$time) pass through the temporal convolution layer with a receptive field of 1$\times$50 sizes and a spatial convolution layer with a 64$\times$1 receptive field. In order to solve the covariate shift problem, batch normalization was used after each convolution layer. Next, average pooling layers were set to resize the convolution, which had a 1$\times$8 kernel size with a stride of 1$\times$8. Then, another convolution layer and average pooling layer with a 1$\times$8 receptive field were applied. Successively, single-layered BiLSTM with 100 hidden states and an average pooling layer were used. Finally, output features were flattened and passed through the fully connected layer with a logarithmic activation function.
We performed 200 iterations for the model training process and saved the model weights and hyper-parameters that produced the lowest loss of the test data. 

\begin{table}[!]
\caption{Classification accuracy of the TINN with 5 repeated test}
\resizebox{\columnwidth}{!}
{%
\begin{tabular}{@{}ccccccc@{}}
\hline
\multirow{2.5}{*}{Subjects} & \multicolumn{6}{c}{Classification accuracy}                         \\ \cline{2-7} 
                          & $1^{st}$ run  & $2^{nd}$ run  & $3^{rd}$ run  & $4^{th}$ run  & $5^{th}$ run  & Average \\ \hline
S1                        & 0.98 & 0.96 & 0.93 & 0.98 & 0.96 & 0.96     \\
S2                        & 0.96 & 0.93 & 0.95 & 0.98 & 0.93 & 0.95     \\
S3                        & 0.86 & 0.90 & 0.90 & 0.87 & 0.89 & 0.88     \\
S4                        & 0.90 & 0.91 & 0.91 & 0.90 & 0.91 & 0.91     \\
S5                        & 0.96 & 0.97 & 0.96 & 0.97 & 0.94 & 0.96     \\
S6                        & 0.95 & 0.94 & 0.95 & 0.96 & 0.96 & 0.95     \\
S7                        & 0.97 & 0.94 & 0.97 & 0.96 & 0.96 & 0.96     \\
S8                        & 0.95 & 0.93 & 0.92 & 0.95 & 0.90 & 0.93     \\
S9                        & 0.85 & 0.89 & 0.92 & 0.86 & 0.86 & 0.88     \\
S10                       & 0.92 & 0.93 & 0.94 & 0.93 & 0.95 & 0.93     \\ \hline
Average             & 0.93 & 0.93 & 0.93 & 0.94 & 0.92 & \textbf{0.93}     \\ \hline
Std.        & 0.04  & 0.02  & 0.02  & 0.04  & 0.03  & \textbf{0.03}      \\ \hline
\end{tabular}
}
\end{table}
\begin{table}[!]
\caption{Decoding performance comparison with the FBCSP, DeepConvNet, ShallowConvNet, EEGNet, and TINN}
\Huge
\resizebox{\columnwidth}{!}
{%
\begin{tabular}{@{}cccccc@{}}
\toprule
\multirow{2}{*}{Subjects} &       & \multicolumn{4}{c}{Methods}                    \\ \cmidrule(l){2-6} 
                          & FBCSP\cite{ang2008filter} & DeepConvNet\cite{schirrmeister2017deep} & ShallowConvNet\cite{schirrmeister2017deep} & EEGNet\cite{lawhern2018eegnet} & TINN \\ \midrule
S1                        & 0.88  & 0.91   & 0.97        & 0.88           & 0.96            \\
S2                        & 0.67  & 0.86   & 0.88        & 0.72           & 0.95            \\
S3                        & 0.59  & 0.82   & 0.81        & 0.73           & 0.88            \\
S4                        & 0.71  & 0.71   & 0.90        & 0.69           & 0.91            \\
S5                        & 0.74  & 0.89   & 0.85        & 0.78           & 0.96            \\
S6                        & 0.70  & 0.95   & 0.92        & 0.74           & 0.95            \\
S7                        & 0.77  & 0.93   & 0.91        & 0.89           & 0.96            \\
S8                        & 0.79  & 0.72   & 0.87        & 0.71           & 0.93            \\
S9                        & 0.87  & 0.82   & 0.88        & 0.73           & 0.88            \\
S10                       & 0.67  & 0.81   & 0.88        & 0.84           & 0.93            \\ \midrule
Average              & 0.74  & 0.84   & 0.89        & 0.77           & \textbf{0.93}   \\ \midrule
Std.                      & 0.09  & 0.08   & 0.04        & 0.07           & \textbf{0.03}   \\ \bottomrule
\end{tabular}}
\end{table}

\section{RESULTS AND DISCUSSION}

To examine different aspects of endogenous paradigms, we calculated the spectral power in six frequency bands ($\delta$-, $\theta$-, $\alpha$-, $\beta$-, low $\gamma$-, and high $\gamma$-band) for each paradigm. Table II shows the ANOVA results for MI, VI, and SI paradigms. Here, we focused on spectral changes depend on paradigms. The differences among the paradigms were statistically significant through all frequency bands (\textit{p}$<$0.001). 
Fig. 2 represents the spatial differences in spectral power between different paradigms and frequency ranges using Bonferroni corrected paired-\textit{t} test. 
The first row of the figure is the result of comparing MI to VI. 
MI had higher activity than VI over the sensorimotor cortex in the $\theta$-band and the motor cortex in the $\alpha$-band. 
VI had higher PSD than MI over the sensorimotor and prefrontal region in the $\delta$-band. Also, the prefrontal and occipital region in the $\alpha$-band had higher PSD, which indicates that higher intensity of brain activation was represented on the prefrontal and visual cortex in the VI tasks\cite{kwon2020decoding}.
The second row is an outcome of comparing MI to SI. In $\delta$-, $\theta$- and high $\gamma$-bands, the MI task had stronger activity over the somatosensory and parietal region than the SI task.
The third row shows the result of comparing VI to SI. In $\gamma$-band, stronger activity was found during the SI task on the left hemisphere, which is a neural pathway from Wernicke's area to Broca's area. Broca's area is involved in the production of coherent speech, and Wernicke's area is involved in speech processing and understanding language\cite{tian2016mental}. In addition, the bilateral activity of the temporal lobe was also found in $\theta$- and $\alpha$-bands. Also, VI had stronger activity over the sensorimotor region in $\delta$- and $\gamma$-bands than the SI task.

\begin{figure}[!]
\centering
\scriptsize
\centerline{\includegraphics[width=\columnwidth]{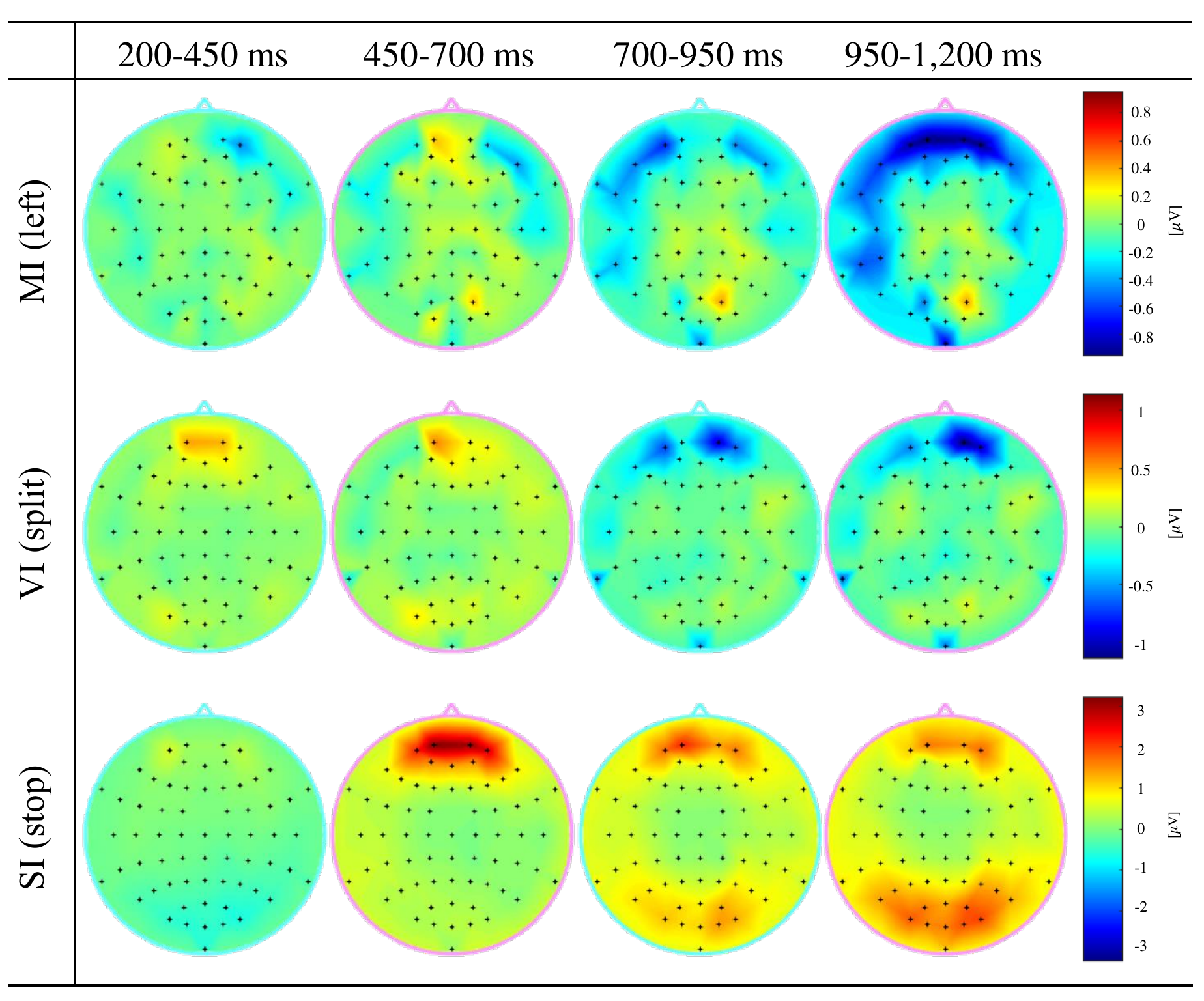}}
\caption{Scalp topographies of the average ERSP for representative classes in each paradigm for the frequency range of 0.5-120 Hz across all subjects.}
\end{figure}

The result shows that the significant frequency bands and brain regions are different for each paradigm. For the effective classification of the three paradigms, all the factors should be included in a common frequency band and channel selection.
Table III represents the result of the statistical analysis on the combinations of representative classes for each endogenous paradigm. Three classes were selected from each of the MI, VI, and SI tasks. In advance, homoscedasticity was satisfied with Levene's test for every class and the Shapiro-Wilk test was satisfied for seven out of nine classes. The classes 'spread out' of VI and 'go' of SI were not satisfied with the normality test. 
The statistical differences among the paradigms were significant except for the groups, which include the classes 'spread-out' and 'go' (\textit{p}$<$0.05).

When we chose the left hand as the class in MI, the \textit{F}-value are the highest regardless of the classes in VI and SI. This result suggests that the spatial features from Broca's and Wernicke's areas on the left hemisphere, which takes part in the linguistic and speech ability\cite{kim1997distinct}, could overlap with the spatial features from the motor cortex on the left hemisphere while right hand MI.

To avoid the overfitting problem, 80$\%$ of the data in each class were randomly selected for the training dataset and the remaining 20$\%$ of data as the test dataset. Therefore, 1,440 epochs (480 trials$\times$3 classes) were used for train dataset and 360 epochs (120 trials$\times$3 classes) were used for test dataset. To evaluate the accuracies using the test dataset in a fair manner, we calculated the accuracies with five repetitions. Table IV shows the classification accuracies of the TINN. The average performance of the TINN was 0.93 ($\pm 0.03$). Table V shows the decoding performance derived by the EEGNet, DeepConvNet, ShallowConvNet, and TINN using the test dataset. The results indicate that the average classification accuracy of the TINN was improved by 0.09 and 0.05 compared to that of DeepConvNet and ShallowConvNet, respectively. In particular, S5 showed the greatest improvement of classification accuracy from 0.85 to 0.96 by using the TINN compared to ShallowConvNet. Moreover, the TINN has the lowest standard deviation for classification accuracies among the subjects, which suggests that our proposed method is more reliable to decode EEG signals than other conventional methods. 

Fig. 3 shows the scalp topographies of the average ERSP for MI, VI, and SI for the frequency range of 0.5-120 Hz. The representative class was selected to visualize the temporal and spatial patterns in each paradigm. Each row of the figure indicates the class within a paradigm, and each column indicates the time window. The time interval of ERSP analysis was set to 200-1,200 ms. In the case of the MI task (left hand), the channel amplitude on the prefrontal and left temporal regions has notably decreased while decreasing the channel amplitude on the prefrontal region only for the VI task (split). Also, the result of SI showed notable spatial and temporal differences in 450-700 ms of time point, the prefrontal region showed high activity, and the occipital region was activated after.

\section{CONCLUSION AND FUTURE WORKS}
Endogenous BMI paradigms such as MI, VI, and SI use the neural signals generated by performing designated mental-imagery tasks without having any restriction on external devices. With the statistical analysis on each paradigm and frequency band, the result shows that the significant frequency bands and brain regions are different for each paradigm. We confirmed that high $\gamma$-band is meaningful frequency bands for SI paradigms, and $\delta$-, $\theta$-, and $\alpha$-bands are useful bands for MI and VI paradigms. Also, we visualized the spatio-temporal features of three different paradigms Through ERSP analysis. Moreover, our proposed TINN outperformed the conventional deep learning methods on classifying various endogenous BMI paradigms with an accuracy of 0.93. This study shows the possibility of multi-class inter-paradigm classification through robust feature extraction from each paradigm.
As future work, multiple subdivision classification on each paradigm and studies on the cross-modal mental imagery tasks would be valuable to expand the limited DoF of the BMI system with robust performances. 
Furthermore, the development of BMI-based human-robot collaboration such as drone swarm control would be an enabling technology to multiply the workforces in various environments.
\addtolength{\textheight}{-7cm}   

\section*{ACKNOWLEDGMENT}
The authors thank to B.-H. Lee, B.-H. Kwon, and G.-H. Shin for their discussion and supports.

\bibliographystyle{IEEEtran}
\bibliography{REFERENCE}

\begin{thebibliography}{10}
\providecommand{\url}[1]{#1}
\csname url@samestyle\endcsname
\providecommand{\newblock}{\relax}
\providecommand{\bibinfo}[2]{#2}
\providecommand{\BIBentrySTDinterwordspacing}{\spaceskip=0pt\relax}
\providecommand{\BIBentryALTinterwordstretchfactor}{4}
\providecommand{\BIBentryALTinterwordspacing}{\spaceskip=\fontdimen2\font plus
\BIBentryALTinterwordstretchfactor\fontdimen3\font minus
  \fontdimen4\font\relax}
\providecommand{\BIBforeignlanguage}[2]{{%
\expandafter\ifx\csname l@#1\endcsname\relax
\typeout{** WARNING: IEEEtran.bst: No hyphenation pattern has been}%
\typeout{** loaded for the language `#1'. Using the pattern for}%
\typeout{** the default language instead.}%
\else
\language=\csname l@#1\endcsname
\fi
#2}}
\providecommand{\BIBdecl}{\relax}
\BIBdecl

\bibitem{lee2020continuous}
D.-H. Lee, J.-H. Jeong, K.~Kim, B.-W. Yu, and S.-W. Lee, ``Continuous {EEG}
  {D}ecoding of {P}ilots’ {M}ental {S}tates {U}sing {M}ultiple {F}eature
  {B}lock-{B}ased {C}onvolutional {N}eural {N}etwork,'' \emph{IEEE Access},
  vol.~8, pp. 121\,929--121\,941, 2020.

\bibitem{prichard2014effects}
G.~Prichard, C.~Weiller, B.~Fritsch, and J.~Reis, ``Effects of different
  electrical brain stimulation protocols on subcomponents of motor skill
  learning,'' \emph{Brain Stimul.}, vol.~7, no.~4, pp. 532--540, 2014.

\bibitem{jantz2017brain}
J.~Jantz, A.~Molnar, and R.~Alcaide, ``A brain-computer interface for extended
  reality interfaces,'' in \emph{ACM SIGGRAPH 2017 VR Village}, 2017, pp. 1--2.

\bibitem{lee2020frontal}
M.~Lee, G.-H. Shin, and S.-W. Lee, ``Frontal {EEG} asymmetry of emotion for the
  same auditory stimulus,'' \emph{IEEE Access}, vol.~8, pp. 107\,200--107\,213,
  2020.

\bibitem{fazel2012p300}
R.~Fazel-Rezai, B.~Z. Allison, C.~Guger, E.~W. Sellers, S.~C. Kleih, and
  A.~K{\"u}bler, ``P300 brain computer interface: current challenges and
  emerging trends,'' \emph{Front. Neuroeng.}, vol.~5, p.~14, 2012.

\bibitem{luck2000event}
S.~J. Luck, G.~F. Woodman, and E.~K. Vogel, ``Event-related potential studies
  of attention,'' \emph{Trends. Cogn. Sci.}, vol.~4, no.~11, pp. 432--440,
  2000.

\bibitem{lee2018high}
M.-H. Lee, J.~Williamson, D.-O. Won, S.~Fazli, and S.-W. Lee, ``A high
  performance spelling system based on {EEG}-{EOG} signals with visual
  feedback,'' \emph{IEEE Trans. Neural Syst. Rehabil. Eng.}, vol.~26, no.~7,
  pp. 1443--1459, 2018.

\bibitem{liu2014recent}
Q.~Liu, K.~Chen, Q.~Ai, and S.~Q. Xie, ``Recent development of signal
  processing algorithms for ssvep-based brain computer interfaces,'' \emph{J.
  Med. Biol. Eng.}, vol.~34, no.~4, pp. 299--309, 2014.

\bibitem{fisher2005photic}
R.~S. Fisher, G.~Harding, G.~Erba, G.~L. Barkley, and A.~Wilkins, ``Photic-and
  pattern-induced seizures: a review for the {E}pilepsy {F}oundation of
  {A}merica {W}orking {G}roup,'' \emph{Epilepsia}, vol.~46, no.~9, pp.
  1426--1441, 2005.

\bibitem{suk2011subject}
H.-I. Suk and S.-W. Lee, ``Subject and class specific frequency bands selection
  for multiclass motor imagery classification,'' \emph{Int. J. Imaging Syst.
  Technol.}, vol.~21, no.~2, pp. 123--130, 2011.

\bibitem{buerkle2021eeg}
A.~Buerkle, W.~Eaton, N.~Lohse, T.~Bamber, and P.~Ferreira, ``{EEG} based arm
  movement intention recognition towards enhanced safety in symbiotic
  human-robot collaboration,'' \emph{Robot. Comput.-Integr. Manuf.}, vol.~70,
  p. 102137, 2021.

\bibitem{liu2021brain}
Y.~Liu, M.~Habibnezhad, and H.~Jebelli, ``Brain-computer interface for
  hands-free teleoperation of construction robots,'' \emph{Autom. Constr.},
  vol. 123, p. 103523, 2021.

\bibitem{koizumi2018development}
K.~Koizumi, K.~Ueda, and M.~Nakao, ``Development of a cognitive brain-machine
  interface based on a visual imagery method,'' in \emph{Conf. Proc. IEEE Eng.
  Med. Biol. Soc. (EMBC)}, 2018, pp. 1062--1065.

\bibitem{chholak2019visual}
P.~Chholak, G.~Niso, V.~A. Maksimenko, S.~A. Kurkin, N.~S. Frolov, E.~N.
  Pitsik, A.~E. Hramov, and A.~N. Pisarchik, ``Visual and kinesthetic modes
  affect motor imagery classification in untrained subjects,'' \emph{Sci.
  Rep.}, vol.~9, no.~1, pp. 1--12, 2019.

\bibitem{lee2020classification}
B.-H. Lee, J.-H. Jeong, K.-H. Shim, and S.-W. Lee, ``Classification of
  high-dimensional motor imagery tasks based on an end-to-end role assigned
  convolutional neural network,'' in \emph{IEEE Int. Conf. Acoust. Speech Sig.
  Proc. (ICASSP)}, 2020, pp. 1359--1363.

\bibitem{schirrmeister2017deep}
R.~T. Schirrmeister, J.~T. Springenberg, L.~D.~J. Fiederer, M.~Glasstetter,
  K.~Eggensperger, M.~Tangermann, F.~Hutter, W.~Burgard, and T.~Ball, ``Deep
  learning with convolutional neural networks for {EEG} decoding and
  visualization,'' \emph{Hum. Brain Mapp.}, vol.~38, no.~11, pp. 5391--5420,
  2017.

\bibitem{ang2008filter}
K.~K. Ang, Z.~Y. Chin, H.~Zhang, and C.~Guan, ``Filter bank common spatial
  pattern ({FBCSP}) in brain-computer interface,'' in \emph{Conf. Proc. IEEE
  Int. Neural Netw. (IJCNN)}, 2008, pp. 2390--2397.

\bibitem{lawhern2018eegnet}
V.~J. Lawhern, A.~J. Solon, N.~R. Waytowich, S.~M. Gordon, C.~P. Hung, and
  B.~J. Lance, ``E{EGN}et: {A} compact convolutional neural network for
  {EEG}-based brain-computer interfaces,'' \emph{J. Neural Eng.}, vol.~15,
  no.~5, p. 056013, 2018.

\bibitem{jeong2020decoding}
J.-H. Jeong, N.-S. Kwak, C.~Guan, and S.-W. Lee, ``Decoding movement-related
  cortical potentials based on subject-dependent and section-wise spectral
  filtering,'' \emph{IEEE Trans. Neural Syst. Rehabil. Eng.}, vol.~28, no.~3,
  pp. 687--698, 2020.

\bibitem{yeung2004neural}
N.~Yeung, M.~M. Botvinick, and J.~D. Cohen, ``The {N}eural {B}asis of {E}rror
  {D}etection: {C}onflict {M}onitoring and the {E}rror-{R}elated
  {N}egativity$.$,'' \emph{Psychol. Rev.}, vol. 111, no.~4, p. 931, 2004.

\bibitem{lee2020design}
D.-H. Lee, H.-J. Ahn, J.-H. Jeong, and S.-W. Lee, ``Design of an {EEG}-based
  {D}rone {S}warm {C}ontrol {S}ystem using {E}ndogenous {BCI} {P}aradigms,''
  \emph{arXiv preprint arXiv:2012.03507}, 2020.

\bibitem{blankertz2010berlin}
B.~Blankertz, M.~Tangermann, C.~Vidaurre, S.~Fazli, C.~Sannelli, S.~Haufe,
  C.~Maeder, L.~E. Ramsey, I.~Sturm, G.~Curio \emph{et~al.}, ``The {B}erlin
  brain--computer interface: {N}on-medical uses of {BCI} technology,''
  \emph{Front. Neurosci.}, vol.~4, p. 198, 2010.

\bibitem{delorme2004eeglab}
A.~Delorme and S.~Makeig, ``{EEGLAB}: {A}n open source toolbox for analysis of
  single-trial {EEG} dynamics including independent component analysis,''
  \emph{J. Neurosci. Methods}, vol. 134, no.~1, pp. 9--21, 2004.

\bibitem{kwon2020decoding}
B.-H. Kwon, J.-H. Jeong, J.-H. Cho, and S.-W. Lee, ``Decoding of {I}ntuitive
  {V}isual {M}otion {I}magery {U}sing {C}onvolutional {N}eural {N}etwork under
  3{D}-{BCI} {T}raining {E}nvironment,'' in \emph{Conf. Proc. IEEE Int. Conf.
  Syst. Man Cybern. (SMC)}, 2020, pp. 2966--2971.

\bibitem{tian2016mental}
X.~Tian, J.~M. Zarate, and D.~Poeppel, ``Mental imagery of speech implicates
  two mechanisms of perceptual reactivation,'' \emph{Cortex}, vol.~77, pp.
  1--12, 2016.

\bibitem{kim1997distinct}
K.~H. Kim, N.~R. Relkin, K.-M. Lee, and J.~Hirsch, ``Distinct cortical areas
  associated with native and second languages,'' \emph{Nature}, vol. 388, no.
  6638, pp. 171--174, 1997.

\end{thebibliography}

\end{document}